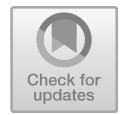

# Multi-task Neural Networks with Spatial Activation for Retinal Vessel Segmentation and Artery/Vein Classification

Wenao Ma[1,2], Shuang Yu[1(✉)], Kai Ma[1], Jiexiang Wang[1,2], Xinghao Ding[2], and Yefeng Zheng[1]

[1] Youtu Lab, Tencent, Shenzhen, China
shirlyyu@tencent.com
[2] School of Information Science and Engineering, Xiamen University, Xiamen, China

**Abstract.** Retinal artery/vein (A/V) classification plays a critical role in the clinical biomarker study of how various systemic and cardiovascular diseases affect the retinal vessels. Conventional methods of automated A/V classification are generally complicated and heavily depend on the accurate vessel segmentation. In this paper, we propose a multi-task deep neural network with spatial activation mechanism that is able to segment full retinal vessel, artery and vein simultaneously, without the pre-requirement of vessel segmentation. The input module of the network integrates the domain knowledge of widely used retinal preprocessing and vessel enhancement techniques. We specially customize the output block of the network with a spatial activation mechanism, which takes advantage of a relatively easier task of vessel segmentation and exploits it to boost the performance of A/V classification. In addition, deep supervision is introduced to the network to assist the low level layers to extract more semantic information. The proposed network achieves pixel-wise accuracy of 95.70% for vessel segmentation, and A/V classification accuracy of 94.50%, which is the state-of-the-art performance for both tasks on the AV-DRIVE dataset. Furthermore, we have also tested the model performance on INSPIRE-AVR dataset, which achieves a skeletal A/V classification accuracy of 91.6%.

## 1 Introduction

Many systemic and cardiovascular diseases have manifestations in the retinal vessels and affect the arteries and veins (A/V) differently [1]. For example, it has been reported that the asymmetrical change of retinal A/V is associated with several cardiovascular diseases [15]. In addition, clinical research has also found that the narrowing of retinal arteriolar caliber is related to the risk of hypertension [3]. Therefore, there is a significant clinical interest in the accurate and automatic A/V classification.

Conventionally, the automated A/V classification in the related literature was performed in a two-stage approach [4,5,11,17,19]. Retinal vessels were first segmented from the background, based on which, vessels were further classified into A/V, using pure hand-crafted features or incorporating the connection information with a graph-based method. Representative works of using a pure feature-based method include [11,17], which extracted the handcrafted features from vessel centerlines and then classified each pixel into artery or vein. Graph based research [4,5,19], on the other hand, first reconstructed the vascular graph from vessel centerline via node analysis or graph estimation, and then classified individual graph trees into arteries or veins.

The above mentioned two-step methods share the same limitation that the performance of A/V classification heavily depends on the accurate vessel segmentation, especially for the graph reconstruction methods. Defects in the vessel segmentation, e.g., broken or wrongly segmented vessels, will be propagated to the subsequent graph reconstruction step, and further influence the performance of A/V classification. Similarly, pure feature-based methods heavily rely on the complex hand-crafted features extracted from the vessel centerline.

Very recently, there has been several emerging works using a Fully Convolutional Network (FCN) to segment and classify retinal A/V at the same time. AlBadawi and Fraz [2] adopted the FCN with an encoder-decoder structure for the pixel-wise classification of A/V. Meyer *et al.* [9] also used FCN for the task of A/V classification and reported performance on major vessels thicker than three pixels. The deep learning based methods have demonstrated the potential to segment A/V in an end-to-end approach. However, the overall vessel segmentation performance of the existing methods using direct A/V segmentation is affected when A/V pixels are classified as background. There is still a research gap of how to improve the vessel segmentation performance together with A/V classification, especially for the capillary vessels.

In this paper, we propose a deep network with a spatial activation mechanism that performs full vessel segmentation and A/V classification simultaneously, so as to improve the accuracy of both tasks. In particular, we design a multi-task output block with an activation mechanism which utilizes the result of vessel segmentation, a relatively easier task, to enhance that of the A/V classification, especially for capillary vessels. The input module of the network integrates the domain knowledge of widely used retinal image processing and filter-based vessel enhancement techniques. In addition, deep supervision modules are attached to the early stages of the encoder section, which can assist the low-level features to extract more semantic information. The proposed framework achieves state-of-the-art performance for both A/V classification and vessel segmentation tasks on AV-DRIVE database, and skeletal A/V classification for INSPIRE-AVR dataset.

## 2   Method

The system workflow of the proposed algorithm is described as below: the color fundus image is firstly processed through modules of illumination correction (IC) and vessel enhancement (VE). Then, patches extracted from those three different sources (original, IC and VE processed) are fed to the proposed deep learning architecture, which further generates three segmentation maps of artery, vein and full vessels simultaneously. The final segmentation and classification maps are generated by stitching the corresponding patches together.

### 2.1   Multi-input Module

The multi-input module of the proposed method integrates the domain knowledge of how the vessel segmentation task is performed conventionally in the non deep learning based methods. In order to remove the non-uniformly distributed brightness across the image, the illumination correction is generally adopted as the pre-processing step for retinal images. In addition, we enhance the vessel map using two conventional vessel segmentation techniques, e.g. the multi-scale Gabor filtering [14] and line detector [10], as auxiliary inputs to the network.

### 2.2   Network Architecture with Spatial Activation

The network architecture of the proposed framework is shown on Fig. 1. We adopt the classic U-Net architecture initialized by a pretrained ResNet as the encoder. In order to accommodate the multi-input channels, we insert an expanding-compressing layer before the ResNet, which first expands the input to a high-dimension space and then compresses the feature maps to three channels, so as to match the input channel number of the ResNet model.

We specifically design a multi-task output block for the effective full vessel segmentation and A/V classification simultaneously. In order to get more accurate results for A/V classification, the network is expected to learn more discriminative features between artery and vein. However, if the network focuses only on A/V classification, it may fail to segment finer capillary vessels. Therefore, it is necessary for the network to learn more common features between A/V, i.e., vessel features. In our proposed framework, we design two parallel branches at the end of the network. One branch focuses on extracting common features between A/V, and generates the probability map of vessel segmentation. Meanwhile, the other one focuses on the discriminative A/V features. Output feature maps of the two branches are then concatenated and further used to generate the final result of A/V classification.

In order to utilize the result of a relatively easier task, i.e. the vessel segmentation in this case, to facilitate the performance of a more complicated task of A/V classification, especially for the capillary vessels, a customized activation block is proposed, as in Eq. (1):

$$m(x) = \sigma(e^{-(x-0.5)^2} - e^{-\frac{1}{4}}) + 1, \qquad (1)$$

where $\sigma$ is the activation factor and set as one in our network. The activation block is designed based on the observation that capillary vessels and boundary pixels generally have a value around 0.5 in the obtained vessel probability map generated by our algorithm, while thick vessels and background pixels have a value near 1 or 0. In order to emphasize the importance of capillary vessels, we adopt a Gaussian function to enhance the weights of pixel values around 0.5. In addition, a bias is added to the activation function to constrain the weight within a range between $[1, 1 + \sigma(1 - e^{-\frac{1}{4}})]$. Then, the activation map is used to adjust each feature map spatially for the A/V classification task, by assigning higher weights for capillary vessel pixels (close to $1 + \sigma(1 - e^{-\frac{1}{4}})$), and lower weights for background and thick vessel pixels (close to 1). In other words, potential capillary vessels can be activated through this process. An example of the activation map can be seen in Fig. 2(E).

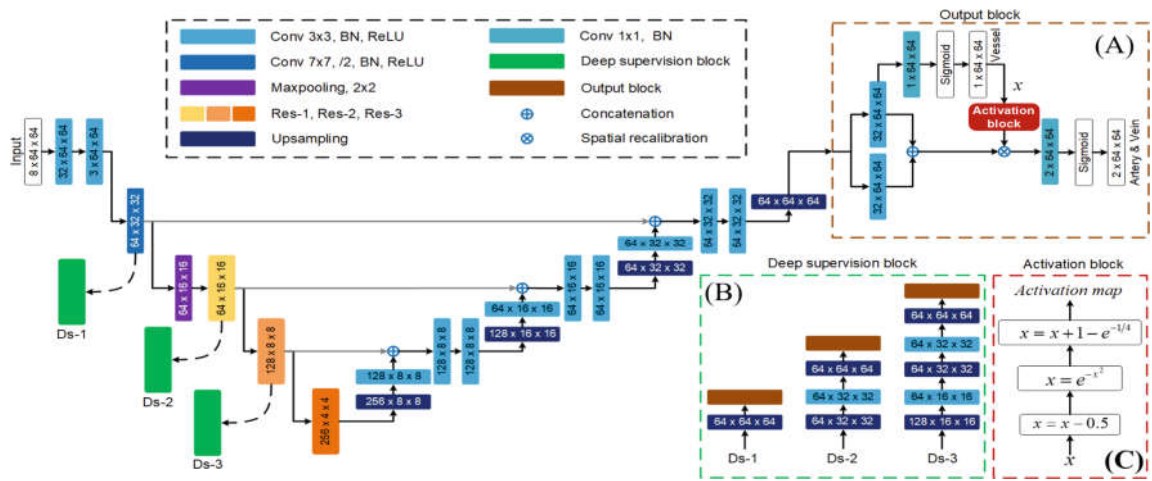

**Fig. 1.** Architecture of the proposed network. (A) Multi-task output block; (B) Deep supervision block; (C) Spatial activation block.

### 2.3 Deep Supervision

As pointed out by Zhang *et al.* [18], a simple fusion of low-level and high-level features, as generally used in U-Net, could be less effective, as there is a gap between the two features in terms of semantic level and spatial resolution. Thus, learning more "semantic" low-level features can help the network to achieve better performance for the U-Net architecture. In addition, the presence of vanishing gradients also makes the loss back-propagation less effective for layers close to the input layer. Considering the two limitations of the network, we introduce deep supervision to the architecture by adding extra side output layers at the encoder section after each ResNet block, as shown in Fig. 1, so as to assist the low-level layers to extract more semantic features and accelerate the convergence.

We also specifically design the loss function accordingly, which contains three elements, including binary cross-entropy loss of the final output, losses of deep supervision blocks and a weight decay regularization term, as in Eqs. (2) and (3):

$$Loss = BCE(output, GT) + \frac{1}{3}\sum_{i=1}^{3} BCE(side_i, GT) + \frac{\lambda}{2}||\Theta||_2^2 \quad (2)$$

$$BCE(pred, target) = -\sum_{c=1}^{3} \mu_c \cdot target_c \cdot \log(pred_c) \quad (3)$$

where $\Theta$ represents the network parameters; $i$ represents the $i^{th}$ deep supervision block; $c$ denotes the $c^{th}$ class of the output; the weight of each class is denoted as $\mu_c$ with $\frac{3}{7}$, $\frac{2}{7}$ and $\frac{2}{7}$ for vessel, artery and vein, respectively.

## 3  Experimental Results

Our model was primarily trained and evaluated on the publicly available AV-DRIVE database [7]. The AV-DRIVE database contains 20 training and 20 test retinal color fundus images with dimension of 584 × 565 pixels, with pixel-wise labeling of vessel segmentation and A/V classification provided. Apart from this, we have also evaluated the model performance on INSPIRE-AVR dataset [12], which contains 40 color images with dimension of 2048 × 2392. Since INSPIRE-AVR provides only A/V labels for centerline, without pixel-wise vessel segmentation, thus it cannot be used to train the model. In order to enrich the usable training data for the model, we have also manually labeled the A/V classification for publicly available High Resolution Fundus (HRF) dataset [13], where the original database contains only pixel-wise vessel segmentation.

In the training stage, patches with size of 64 × 64 were randomly extracted from the retinal images and fed to the network. Whereas in the test stage, ordered patches were extracted at the stride of 10 and final result was obtained by stitching the corresponding patch predictions together. Stochastic gradient descent with momentum was adopted to optimize the model for a maximum of 60,000 iterations with batch size of 16. The initial learning rate was set as 0.05 and halved every 7500 iterations. The training process tooks around 2 h on a NVIDIA Tesla P40 GPU and it takes around 8 s to segment one image during the test phase.

Figure 2 shows a representative performance for vessel segmentation and A/V classification result for an image from the AV-DRIVE test set. Note the enlarged view of two representative patches deemed as challenging for the conventional methods, with the upper one being complex crossovers and lower one being two close-by parallel vessels. Under the multi-task framework, our model is able to accurately segment the overlapping vessels and classify the complex vessels.

### 3.1  Ablation Studies

Comprehensive ablation studies have been performed to evaluate the contribution of different modules of the proposed model, including the multi-task (MTs),

Multi-input (MIs) and spatial activation mechanism (AC). The baseline model was built by removing the above mentioned three modules and direct segmenting A/V pixels from the background. We adopt four metrics for the evaluation of vessel segmentation: the average accuracy (Acc), sensitivity (Sen), specificity (Sp), and area under curve (AUC). A/V classification performance is evaluated using pixel-wise Acc, Sen and Sp for the ground-truth artery vein pixels. By taking arteries as positives and veins as negatives, Sen reflects the model's capability of detecting arteries and Sp for veins.

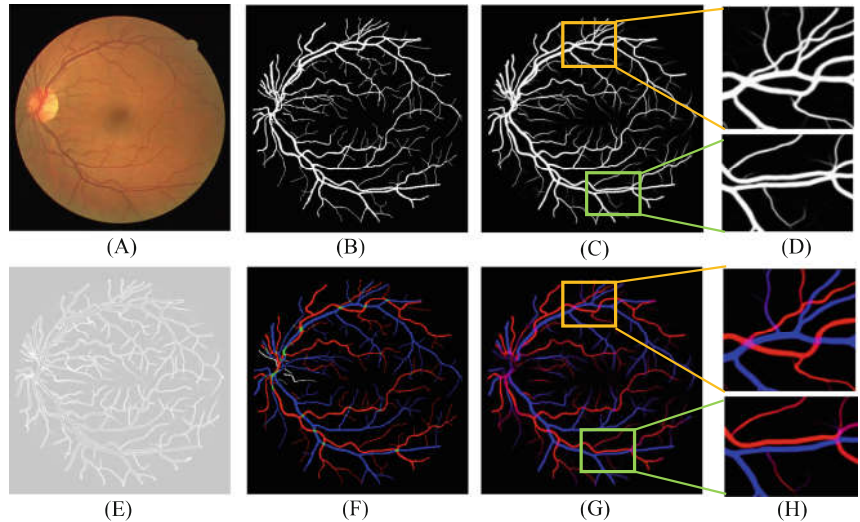

**Fig. 2.** Representative performance of vessel segmentation and A/V classification result. (A) original image from the AV-DRIVE test set; (E) activation map; (B)(F) manual label of vessel segmentation and A/V classification; (C)(G) model segmentation result; (D)(H) enlarged view of two representative challenging regions for the conventional method.

**Table 1.** The ablation study results of vessel segmentation and A/V classification.

| Combination | | | Vessel segmentation | | | | A/V classification | | |
|---|---|---|---|---|---|---|---|---|---|
| MTs | MIs | AC | Acc(%) | Sen(%) | Sp(%) | AUC(%) | Acc(%) | Sen(%) | Sp(%) |
|  |  |  | 94.98 | 68.86 | **98.79** | 97.60 | 91.25 | 89.68 | 92.55 |
| ✓ |  |  | 95.61 | 78.50 | 98.10 | 98.01 | 91.63 | 90.46 | 92.63 |
| ✓ | ✓ |  | 95.66 | 78.30 | 98.19 | 98.08 | 91.98 | 90.36 | **93.42** |
| ✓ | ✓ | ✓ | **95.70** | **79.16** | 98.11 | **98.10** | **92.58** | **92.18** | 92.98 |

As listed in Table 1, by using the multi-task module in the model, both vessel segmentation and A/V classification performances are improved, by 0.6% and 0.4% respectively. This indicates the necessity of a multi-task solution to the problem, which improves the performance for both tasks. The integration of domain knowledge from conventional methods increases the A/V classification performance by 0.3%. In particular, when we add the activation mechanism to the network, the A/V classification performance is further boosted by 0.6%. Compared with A/V classification, the vessel segmentation task is a relatively easier task. The special design of the multi-task out block and activation mechanism takes advantage of the performance of the easier task to enhance that of a more complicated one. And ablation study has proved the effectiveness of this approach. At last, the proposed network achieves a pixel-wise accuracy of 95.70% for vessel segmentation and 92.58% for A/V classification.

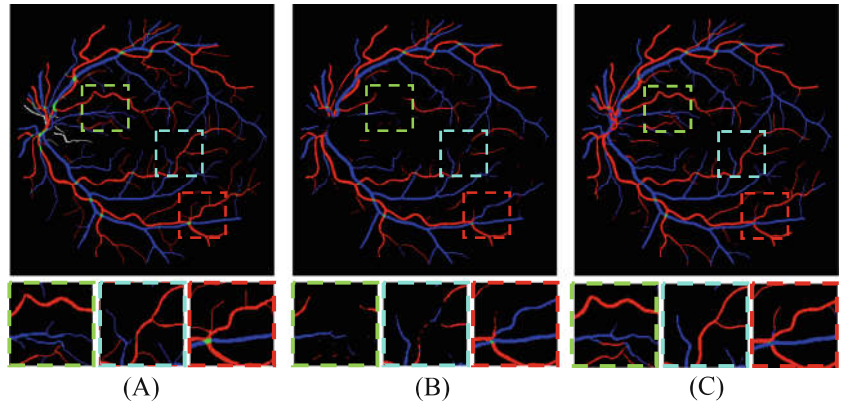

**Fig. 3.** Comparison of model performance for baseline and proposed method. (A) Ground truth; (B) Baseline model performance; (C) Proposed method.

Figure 3 visualizes the performances of the proposed model against that of the baseline. As marked on the enlarged views, the proposed method has remarkably improved the segmentation and classification of A/Vs over the baseline. More capillary vessels are segmented and correctly classified by the proposed method where the baseline model fails to.

### 3.2 Comparison to Existing Methods

Tables 2 and 3 list the performance comparison between existing methods and our proposed model on the AV-DRIVE database. Note in Table 3, existing methods evaluated the A/V classification performance on the segmented vessels only. In contrast, we evaluate the performance on all the ground truth artery/vein pixels, which is relatively more strict than that on the segmented vessels, since the classification of major vessels is comparatively an easier task if capillary vessels are not segmented. When evaluating under the same criteria with existing methods, our model achieves a pixel-wise accuracy of 94.50%, which surpasses the current best A/V classification method by a noticeable margin.

In addition, we have tested our model on the INSPIRE-AVR dataset. The training set contains 20 images from AV-DRIVE and 45 images from HRF dataset, which contains publicly available vessel segmentation label and we manually labeled the A/V class. The skeletal A/V classification of INSPIRE-AVR achieved 91.6% without fine-tuning, which is the state-of-the-art performance.

**Table 2.** Performance comparison of vessel segmentation on the AV-DRIVE dataset.

| Methods | Acc(%) | Sp(%) | Sen(%) | AUC(%) |
|---|---|---|---|---|
| Fu *et al.* [6] | 94.70 | - | 72.94 | - |
| Liskowski *et al.* [8] | 95.35 | 98.07 | 78.11 | 97.90 |
| Wu *et al.* [16] | 95.67 | **98.19** | 78.44 | 98.07 |
| **Proposed** | **95.70** | 98.11 | **79.16** | **98.10** |

**Table 3.** Performance comparison of A/V classification on AV-DRIVE and INSPIRE-AVR datasets.

| Methods | AV-DRIVE | | | INSPIRE | | |
|---|---|---|---|---|---|---|
|  | Acc(%) | Sen(%) | Sp(%) | Acc(%) | Sen(%) | Sp(%) |
| Dashtbozorg *et al.* [4] | 87.4 | 90.0 | 84.0 | 84.9 | 91.0 | 86.0 |
| Estrada *et al.* [5] | 93.5 | 93.0 | 94.1 | 90.9 | 91.5 | 90.2 |
| Xu *et al.* [17] | 92.3 | 92.9 | 91.5 | - | - | - |
| Zhao *et al.* [19] | - | 91.9 | 91.5 | 91.0 | 91.8 | 90.2 |
| **Proposed (GT)** | 92.6 | 92.2 | 93.0 | 90.3 | 91.4 | 89.7 |
| **Proposed** | **94.5** | **93.4** | **95.5** | **91.6** | **92.4** | **91.3** |

## 4 Conclusion

In this paper, we proposed a novel multi-tasking neural network with spatial activation mechanism that enables the end-to-end segmentation of artery, vein and full vessel maps simultaneously. We evaluated our method on the AV-DRIVE dataset and compared it to the other existing research. The result shows that our method outperforms the existing methods, achieving state-of-the-art performance for both vessel segmentation and A/V classification tasks.

The proposed framework has significantly improved the accuracy and efficiency of A/V classification, which lays the foundation for quantitative vessel parameterization. In the near future, fully automatic vascular parameter generation modules will be implemented and validated on the basis of the current work, so as to facilitate the clinical retinal vascular biomarker study.

**Acknowledgment.** This work was funded by the Key Area Research and Development Program of Guangdong Province, China (No. 2018B010111001).

Multi-task NN for Retinal Vessel Seg and A/V Classification     7773. Chew, S.K., Xie, J., Wang, J.J.: Retinal arteriolar diameter and the prevalence and incidence of hypertension: a systematic review and meta-analysis of their association. Curr. Hypertens. Rep. **14**(2), 144–151 (2012)
4. Dashtbozorg, B., Mendonça, A.M., Campilho, A.: An automatic graph-based approach for artery/vein classification in retinal images. IEEE Trans. Image Process. **23**(3), 1073–1083 (2014)
5. Estrada, R., Allingham, M.J., Mettu, P.S., Cousins, S.W., Tomasi, C., Farsiu, S.: Retinal artery-vein classification via topology estimation. IEEE Trans. Med. Imaging **34**(12), 2518–2534 (2015)
6. Fu, H., Xu, Y., Wong, D.W.K., Liu, J.: Retinal vessel segmentation via deep learning network and fully-connected conditional random fields. In: 2016 IEEE 13th International Symposium on Biomedical Imaging (ISBI), pp. 698–701. IEEE (2016)
7. Hu, Q., Abràmoff, M.D., Garvin, M.K.: Automated separation of binary overlapping trees in low-contrast color retinal images. In: Mori, K., Sakuma, I., Sato, Y., Barillot, C., Navab, N. (eds.) MICCAI 2013. LNCS, vol. 8150, pp. 436–443. Springer, Heidelberg (2013). https://doi.org/10.1007/978-3-642-40763-5_54
8. Liskowski, P., Krawiec, K.: Segmenting retinal blood vessels with deep neural networks. IEEE Trans. Med. Imaging **35**(11), 2369–2380 (2016)
9. Meyer, M.I., Galdran, A., Costa, P., Mendonça, A.M., Campilho, A.: Deep Convolutional artery/vein classification of retinal vessels. In: Campilho, A., Karray, F., ter Haar Romeny, B. (eds.) ICIAR 2018. LNCS, vol. 10882, pp. 622–630. Springer, Cham (2018). https://doi.org/10.1007/978-3-319-93000-8_71
10. Nguyen, U.T., Bhuiyan, A., Park, L.A., Ramamohanarao, K.: An effective retinal blood vessel segmentation method using multi-scale line detection. Pattern Recognit. **46**(3), 703–715 (2013)
11. Niemeijer, M., van Ginneken, B., Abràmoff, M.D.: Automatic classification of retinal vessels into arteries and veins. In: Medical Imaging 2009: Computer-Aided Diagnosis, vol. 7260, p. 72601F (2009)
12. Niemeijer, M., et al.: Automated measurement of the arteriolar-to-venular width ratio in digital color fundus photographs. IEEE Trans. Med. Imaging **30**(11), 1941–1950 (2011)
13. Odstrcilik, J., et al.: Retinal vessel segmentation by improved matched filtering: evaluation on a new high-resolution fundus image database. IET Image Process. **7**(4), 373–383 (2013)
14. Soares, J.V.B., Leandro, J.J.G., Cesar, R.M., Jelinek, H.F., Cree, M.J.: Retinal vessel segmentation using the 2-d gabor wavelet and supervised classification. IEEE Trans. Med. Imaging **25**(9), 1214–1222 (2006)
15. Wong, T.Y., et al.: Retinal arteriolar narrowing and risk of coronary heart disease in men and women: the atherosclerosis risk in communities study. JAMA **287**(9), 1153–1159 (2002)
16. Wu, Y., Xia, Y., Song, Y., Zhang, Y., Cai, W.: Multiscale network followed network model for retinal vessel segmentation. In: Frangi, A.F., Schnabel, J.A., Davatzikos, C., Alberola-López, C., Fichtinger, G. (eds.) MICCAI 2018. LNCS, vol. 11071, pp. 119–126. Springer, Cham (2018). https://doi.org/10.1007/978-3-030-00934-2_14
17. Xu, X., Ding, W., Abràmoff, M.D., Cao, R.: An improved arteriovenous classification method for the early diagnostics of various diseases in retinal image. Comput. Methods Programs Biomed. **141**, 3–9 (2017)